\documentclass{article}
\usepackage{spconf,amsmath,graphicx}

\usepackage{hyperref}

\title{LARGE-SCALE WEAKLY SUPERVISED AUDIO CLASSIFICATION USING GATED CONVOLUTIONAL NEURAL NETWORK}
\name{Yong Xu*, Qiuqiang Kong*, Wenwu Wang, Mark D. Plumbley\thanks{* These first two authors contributed equally to this work.}}
\address{Center for Vision, Speech and Signal Processing, University of Surrey, UK}

\email{\{yong.xu, q.kong, w.wang, m.plumbley\}@surrey.ac.uk}
\begin{document}
%
\maketitle
\begin{abstract}
In this paper, we present a gated convolutional neural network and a temporal attention-based localization method for audio classification, which won the 1st place in the large-scale weakly supervised sound event detection task of Detection and Classification of Acoustic Scenes and Events (DCASE) 2017 challenge. The audio clips in this task, which are extracted from YouTube videos, are manually labelled with one or a few audio tags but without time stamps of the audio events, which is called as weakly labelled data. Two sub-tasks are defined in this challenge including audio tagging and sound event detection using this weakly labelled data. A convolutional recurrent neural network (CRNN) with learnable gated linear units (GLUs) non-linearity applied on the log Mel spectrogram is proposed.
In addition, a temporal attention method is proposed along the frames to predicate the locations of each audio event in a chunk from the weakly labelled data. We ranked the 1st and the 2nd as a team in these two sub-tasks of DCASE 2017 challenge with F value 55.6\% and Equal error 0.73, respectively. 
\end{abstract}
\section{Introduction}
\label{sec:intro}

Audio classification is a task to classify audio recordings into different classes. Weakly labelled audio data contains only the presence or absence of the audio events but without the time stamps of the audio events \cite{kumar2016audio}. Weakly labelled audio classification has many applications in information retrieval \cite{bogdanov2013essentia}, surveillance of abnormal sound in public area and industry use \cite{dimitrov2014analyzing}. Some challenges divide the audio classification to subtasks including audio scene classification \cite{mesaros2016tut} and sound event detection \cite{mesaros2016tut}. Recently a large-scale weakly supervised sound event detection task
of Detection and Classification of Acoustic Scenes and Events (DCASE) 2017 challenge \cite{mesaros2017dcase} was proposed where the data set is a subset of Google AudioSet \cite{gemmeke2017audio} containing both transportation and warnings sounds. This task includes an audio tagging (AT) \cite{xu2017trans} subtask and a weakly supervised sound event detection (SED) \cite{qq2017icassp} subtask. The AT task aims to predict one or a few labels of an audio recording and SED needs to predict the time stamps of the audio events. 

Many audio classification methods are based on the bag of frames \cite{ye2015acoustic} assumption, where an audio recording is cut into segments and each segment inherits the labels of the audio recording. However this assumption is incorrect because some audio events only happen a short time in an audio clip. Multi-instance learning (MIL) \cite{kumar2016audio} has been applied to train on the weakly labelled data. Recently state-of-the-art audio classification methods \cite{choi2016automatic, parascandolo2017convolutional} transform the waveform to the time-frequency (T-F) representation. Then the T-F representation is treated as an image which is fed into CNNs. However, unlike image classification where the objects are usually centered and occupies a dominant part of the image, audio events may only occur a short time in an audio recording. To solve this problem, some attention models \cite{yongIS2017} for audio classification are applied to attend to the audio events and ignore the background sounds. 

In this paper, we propose a unified neural network model which fits for both of the audio tagging task and the weakly sound event detection task, simultaneously. The first contribution of this paper is to apply the learnable gated linear unit (GLU) \cite{dauphin2016language} to replace the ReLU activation \cite{nair2010rectified} after each layer of the convolutional neural network for audio classification. This learnable gate is able to control the information flow to the next layer. When a gate is close to 1, then the corresponding T-F unit is attended. When a gate is close to 0, then the corresponding T-F unit is ignored.
Following the convolutional layers, the recurrent layer is followed to utilize the temporal information. Then the temporal attention method is proposed to localize the audio events in a chunk. This attention part will attend to the audio events and ignore unrelated audio segments hence it is able to do sound event detection from weakly labeled data.

The paper is organized as follows. Section \ref{sec:at} introduces the gated linear units in the neural network. Section \ref{sec:sed} proposed the localization method for audio events from the weakly labeled data. Section \ref{sec:exp} shows experiments. Section \ref{sec:conclusion} summaries and forecasts the future work. 
\begin{figure*}[htb]

\begin{minipage}[b]{1.0\linewidth}
  \centering
  \centerline{\includegraphics[width=0.95\textwidth]{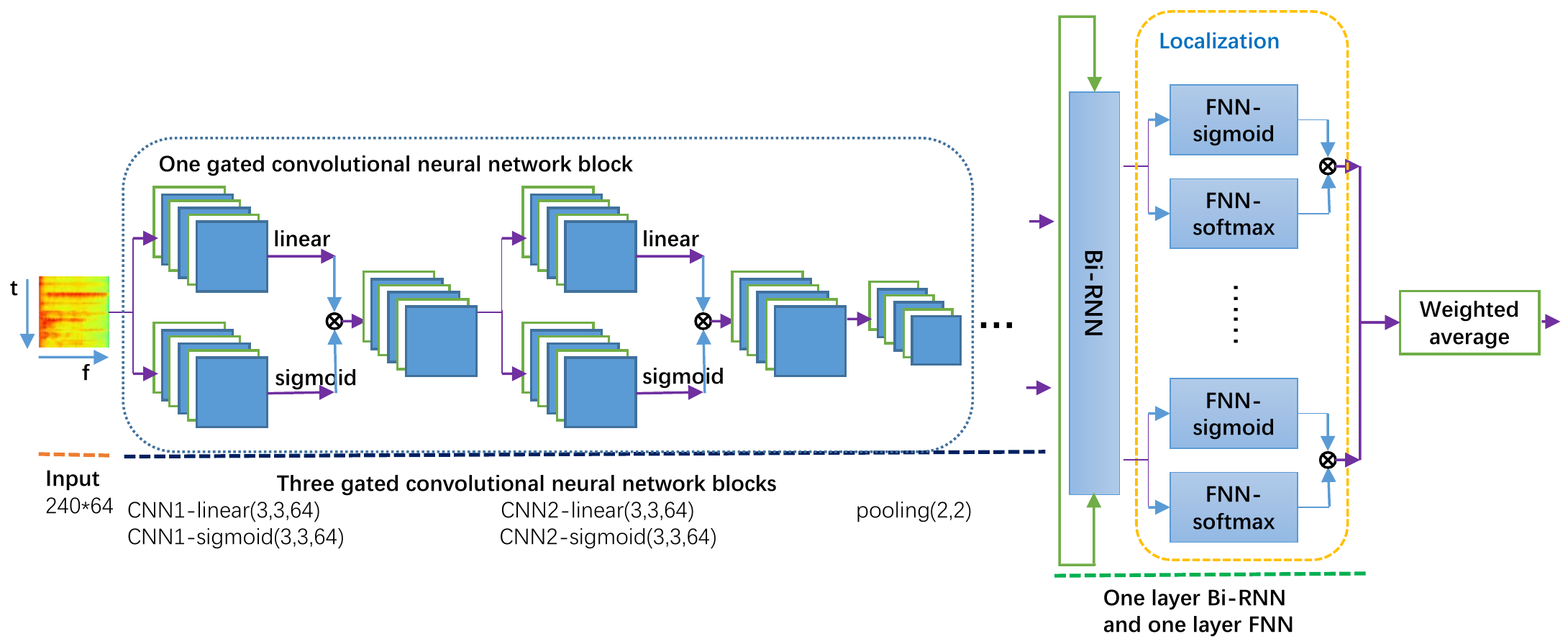}}
\end{minipage}
\caption{The diagram of the proposed unified model for audio tagging and weakly supervised sound event detection (SED). The final outputs are the audio tagging predictions. SED predictions are extracted from the intermediate localization module.}
\label{fig:cgrnn_glu}
\end{figure*}

\section{PROPOSED GATED LINEAR UNITS IN CRNN for AUDIO TAGGING}
\label{sec:at}
In this section, the Convolution recurrent neural networks (CRNNs) baseline, gated linear unit (GLU), mini-batch data balancing and system fusion will be introduced.

\subsection{CRNN baseline}

CRNN has been successfully used in audio classification task \cite{adavanne2017sound, parascandolo2017convolutional}. For the audio tagging task, a CRNN-based method has been proposed in \cite{yongIJCNN2017, yongIS2017} to predict the audio tags. First the waveform of the audio recordings are transformed to T-F representation such as log Mel spectrogram. Then convolutional layers are applied on the T-F representation to extract high level features. Then a bi-directional recurrent neural network (Bi-RNN) are adopted to capture the temporal context information followed by a feed-forward neural network (FNN) with the number of audio classes to predict the posteriors of each audio class at each frame. Finally, the prediction probability of each audio tag is obtained by averaging the posteriors of all the frames. 

In the training phase, we apply the binary cross-entropy loss between the predicted probability and the ground truth of an audio recording. The weights of the neural network can be updated by the gradient of the weights computed using back-propagation. The loss can be defined as:
\begin{equation}
E=-\sum_{n=1}^{N}(\textbf{P}_{n}\text{log}{\textbf{O}}_{n}+(1-\textbf{P}_{n})\text{log}(1-{\textbf{O}}_n))
\label{eq:DNNerrors_bce}
\end{equation}

where $E$ is the binary cross-entropy, ${\textbf{O}}_n$ and $\textbf{P}_{n}$ denote the estimated and reference tag vector at sample index $n$, respectively. The bunch size is represented by $N$. 
Adam \cite{kingma2014adam} is used as the stochastic optimization method.

\subsection{Gated linear units in CNNs}
We propose to use the gated linear units (GLUs) \cite{dauphin2016language} as activation to replace the ReLU \cite{nair2010rectified} activation in the CRNN model. GLUs are first proposed in \cite{dauphin2016language} for language modeling. The motivation of using GLUs in audio classification is to introduce the attention mechanism to all the layers of the neural network. The GLUs can control the amount of information of a T-F unit flow to the next layer. If a GLU is close to 1, then the corresponding T-F unit should be attended. If a GLU is close to 0, then the corresponding T-F unit should be ignored. By this means the network will learn to attend to the audio events and ignore the unrelated sounds. GLUs are defined as:
\begin{equation}
    \textbf{Y}=(\textbf{W}*\textbf{X}+\textbf{b})\odot\sigma(\textbf{V}*\textbf{X}+\textbf{c})
\end{equation}
where $\sigma$ is the sigmoid non-linearity and $\odot$ is the element-wise product and $\ast$ is the convolution operator. $\textbf{W}$ and $\textbf{V}$ are the convolutional filters, $\textbf{b}$ and $\textbf{c}$ are the biases. $ \textbf{X} $ denotes the input T-F representation in the first layer or the feature maps of the interval layers. 

The framework of the model is shown in Fig. \ref{fig:cgrnn_glu}, a pair of convolutional networks are used to generate the gating outputs and the linear outputs. These GLUs can reduce the gradient vanishing problem for deep networks by providing a linear path for the gradients while retaining non-linear capabilities through the sigmoid operation. The output of each layer is a linear projection $(\textbf{W}*\textbf{X}+\textbf{b})$ modulated by the gates $\sigma(\textbf{V}*\textbf{X}+\textbf{c})$. Similar to the gating mechanisms in long short-term memories (LSTMs) \cite{hochreiter1997long} or gated recurrent units (GRUs) \cite{chung2014empirical}, these gates multiply each element of the matrix $(\textbf{W}*\textbf{X}+\textbf{b})$ and control the information passed on in the hierarchy \cite{dauphin2016language}. From the feature selection view, the GLUs can be regarded as an attention scheme on the time-frequency (T-F) bin of each feature map. It can attend to the T-F bin with related audio events by setting its value close to one otherwise close zero.

\subsection{Mini-batch data balancing}
The data set defined in this challenge is highly unbalanced, that means the number of samples of each class varies a lot. For example, the `car' class occurred 25744 times in the data set while `car alarm' only occurred 273 times. This highly unbalanced data will bias the training to the class with a large number of occurrences. As we are using mini-batch to train the network, there is a extreme situation where all the samples in a mini-batch are `car'. To solve this problem we balance the frequency of different classes in a mini-batch to ensure that the number of most frequent samples is at most 5 times than the least frequent samples in a mini-batch averagely. 

\subsection{System results fusion}
System results fusion is important to improve the robustness of systems. In this work, we adopt two level fusion strategies. As neural networks are trained by the stochastic gradient descent (SGD)
algorithm with a fixed or dynamically changing learning rate, the performance will be gradually better but fluctuant along the epochs. Hence, our first fusion strategy is conducted among the epochs in the same system. This will improve its stability of the system. The second fusion strategy is to average the posteriors from different systems with different configurations.

\section{PROPOSED LOCALIZATION FOR WEAKLY SUPERVISED SOUND EVENT DETECTION}
\label{sec:sed}
Different from the audio tagging task without knowing the temporal locations of each audio event which is presented in Sec. \ref{sec:at}, the sound event detection (SED) task needs to predict the temporal locations of each occurring audio event. The problem would be more difficult if there was no strong labels, namely frame-level labels. This is the so-called weakly supervised SED defined in the task 4 of DCASE2017 challenge.



As shown in Fig. \ref{fig:cgrnn_glu}, an additional feed-forward neural network with softmax as the activation function is introduced to help to infer the temporal locations of each occurring class. To keep the time resolution of the input whole audio spectrogram, we adjust the pooling steps in CNNs shown in Fig. \ref{fig:cgrnn_glu} by only pooling on the spectral axis while not pooling on the time axis. So the feed-forward with sigmoid as the activation function shown in Fig. \ref{fig:cgrnn_glu} will do classification at each frame, meanwhile the feed-forward with softmax as the activation function shown in Fig. \ref{fig:cgrnn_glu} will attend on the most salient frames for each class.

If we define the FNN-softmax output, $\textbf{Z}_{\text{loc}}(t)$, as the localization vector, then it is multiplied with the classification output $\textbf{O}(t)$ at each frame as,
\begin{equation}
\textbf{O}^{\prime}(t)=\textbf{O}(t) \odot \textbf{Z}_{\text{loc}}(t)
\label{eq:att_det_tag}
\end{equation}
where $\odot$ represents the element-wise multiplication. To get the final acoustic event tag predictions, $\textbf{O}^{\prime}(t)$ should be averaged across the audio chunk to get the final output $\textbf{O}^{\prime\prime}$. $\textbf{O}^{\prime\prime}$ is defined as the weighted average of $\textbf{O}^{\prime}(t)$ as following,
\begin{equation}
\textbf{O}^{\prime\prime}=\frac{\sum_{t=0}^{T-1}\textbf{O}^{\prime}(t)}{\sum_{t=0}^{T-1}\textbf{Z}_{\text{loc}}(t)}
\label{eq:att_det_tag_final}
\end{equation}
where $T$ is final frame-level resolution along the spectrogram. If there is no pooling along the time axis, $T$ will be the same with the frame number of the whole audio spectrogram considering that the input is the whole spectrogram.

Note that there is no frame-level strong labels, the temporal locations of each occurring class can only be weakly-supervised inferred as intermediate variables. The back-propagate loss is the same with the audio tagging task by comparing the reference labels with the final output $\textbf{O}^{\prime\prime}$. 

\section{Experiments and results} \label{sec:exp}
\subsection{Experimental setup}
The task employs a subset of Google Audioset \cite{audioset}. AudioSet consists of an expanding ontology of 632 sound event classes and a collection of 2 million human-labeled 10-second sound clips (less than 21\% are shorter than 10-seconds) drawn from 2 million YouTube videos. The ontology is specified as a hierarchical graph of event categories, covering a wide range of human and animal sounds, musical instruments and genres, and common everyday environmental sounds. The subset consists of 17 sound events divided into two categories: ``Warning'' and ``Vehicle''.

Log-Mel filter banks and Mel-frequency cepstral coefficients (MFCCs) are used as our features. Each chunk has 240 frames by 64 mel frequency channels. As shown in Fig. \ref{fig:cgrnn_glu}, three gated convolutional neural network blocks are adopted. Each convolutional network has 64 filters with 3*3 size. The pooling size is 2*2 for the audio tagging sub-task while it is 1*2 for the sound event detection sub-task. One bi-directional gated recurrent neural network with 128 units is used. The feed-forward neural network has 17 output nodes where each of them is corresponding to each audio event class. The learning rate is fixed to be 0.001.

The source codes for this paper can be downloaded from Github\footnote{\url{https://github.com/yongxuUSTC/dcase2017_task4_cvssp}}.

\subsection{Results}
In this section, the audio tagging results and then the weakly supervised sound event detection results will be given.
\subsubsection{Audio tagging}

Table \ref{tab:at} presents the F1, Precision and Recall comparisons for the \textbf{audio tagging} sub-task on the development set and the evaluation set. ``CRNN-logMel-noBatchBal'' denotes the CRNN system trained without mini-batch data balancing strategy. The DCASE2017 baseline model was multilayer perceptron (MLP) based method \cite{mesaros2017dcase}. Our proposed CRNN systems show much better performance. Compared the CRNNs with/without mini-batch balancing, data balancing is important to get higher F1, precision and recall scores. The proposed gated CRNN also gains effective improvement. The final fusion system is conducted by combining the system trained on different features, namely log Mel and MFCC. On the evaluation set which is a blinding test, our system ranks 1st in this audio tagging challenge according to the more comprehensive score, namely F1. CNN-ensemble \cite{Lee2017a} and Frame-CNN \cite{Chou2017} ranks 2nd and 3rd as a team, respectively. Note that our system has a notable absolute 3\% improvement over the 2nd system \cite{Lee2017a}.

\begin{table}[h]  
	\centering
	\caption{F1, Precision and Recall comparisons for the \textbf{audio tagging} sub-task on the development the {evaluation sets}.}
\begin{tabular}{lccc}
\hline
\textbf{Dev-set} & \textbf{F1} & \textbf{Precision} & \textbf{Recall} \\
\hline
DCASE2017 Baseline \cite{mesaros2017dcase} & 10.9  & 7.8   & 17.5 \\
\hline
CRNN-logMel-noBatchBal & 42.0  & 47.1  & 38.0 \\
\hline
CRNN-logMel (i) & 52.8  & 49.9  & 56.1 \\
\hline
Gated-CRNN-logMel (ii) & 56.7  & 53.8  & \textbf{60.1} \\
\hline
Gated-CRNN-MFCC (iii) & 52.1  & 51.7  & 52.5 \\
\hline
Fusion (ii+iii) & \textbf{57.7} & \textbf{56.5} & 58.9 \\
\hline
\hline
\textbf{Eval-set} & \textbf{F1} & \textbf{Precision} & \textbf{Recall} \\
\hline
DCASE2017 Baseline \cite{mesaros2017dcase} & 18.2  & 15.0  & 23.1 \\
\hline
CNN-ensemble \cite{Lee2017a} & 52.6  & \textbf{69.7} & 42.3 \\
\hline
Frame-CNN \cite{Chou2017} & 49.0  & 53.8  & 45.0 \\
\hline
Our gated-CRNN-logMel & 54.2  & 58.9  & 50.2 \\
\hline
Our fusion system & \textbf{55.6} & 61.4  & \textbf{50.8} \\
\hline
\end{tabular}%
	\label{tab:at}
\end{table}

\subsubsection{Weakly supervised sound event detection (SED)}
The results of F1 and Error rate comparisons on the development set and the evaluation set are given in Table \ref{tab:sed}. Our proposed gated-CRNN-logMel method outperforms the DCASE2017 baseline \cite{mesaros2017dcase}. With the fusion system, we ranks 2nd as a team in the sound event detection sub-task. The 1st place team achieves 0.66 Error rate and 55.5\% F1 score \cite{Lee2017a}. However, \cite{Lee2017a} used segment input for SED, separately. It assumed that audio events occured everywhere along the chunk. Our method is a unified method without any assumption. Attention based localization seems to be reasonable for weakly supervised SED.

\begin{figure}[t]
	\centering
	\centerline{\includegraphics[width=\columnwidth]{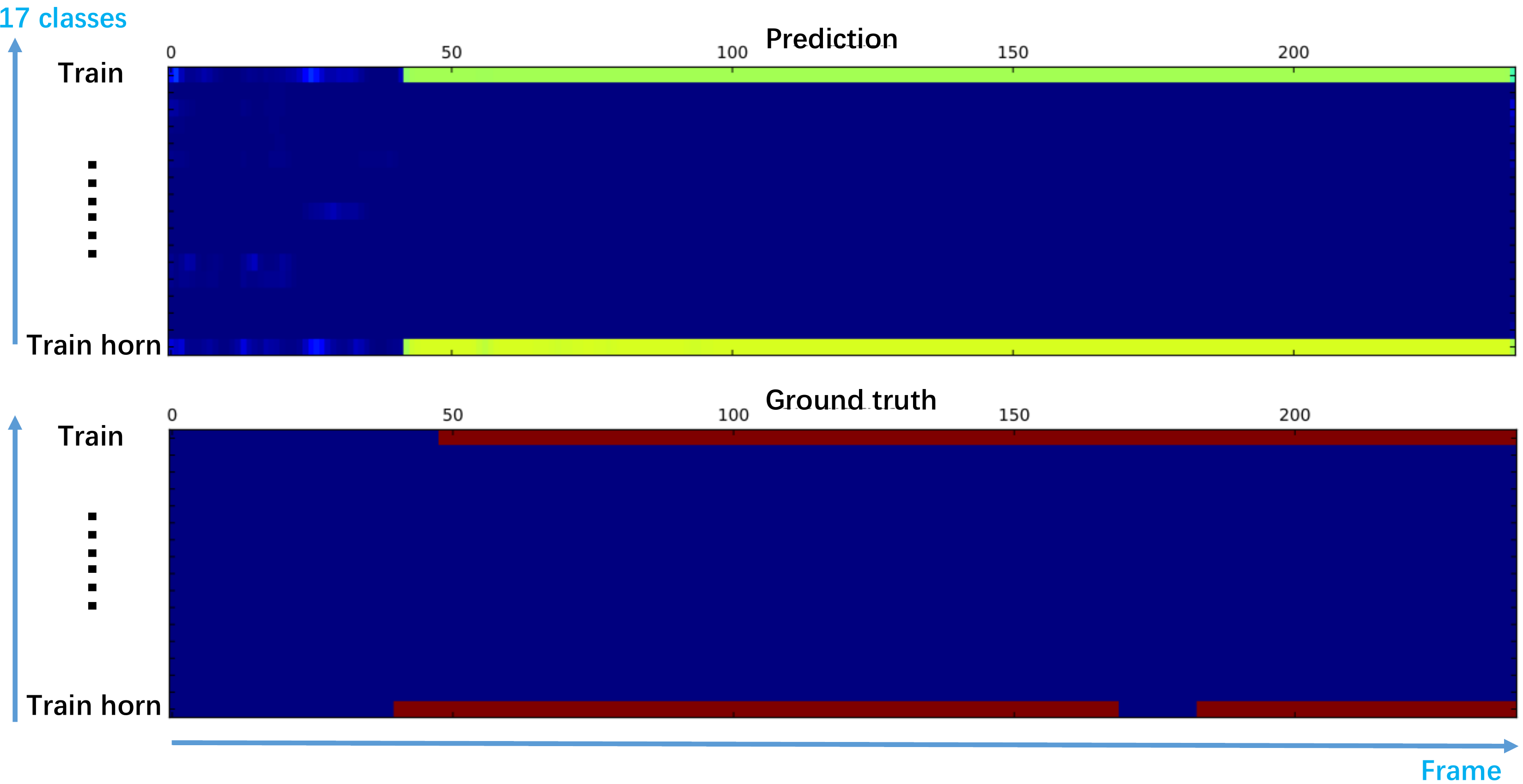}}
	\caption{An example for predicting locations along 240 frames for ``10i60V1RZkQ\_210.000\_220.000.wav'' using the proposed localization method.}
	\label{fig:attention_demo}
\end{figure}

Fig. \ref{fig:attention_demo} shows an example for predicating temporal locations along 240 frames for occurring audio events, namely `train' and `train horn'. Our proposed localization method can almost successfully detect the accurate temporal locations for occurring events, except for the small segment false alarm for the `train horn'.
\begin{table}[h]  
	\centering
	\caption{The results of F1 and Error rate comparisons on the development set and the evaluation set for the \textbf{sound event detection} sub-task are given in Table \ref{tab:sed}. among several methods across the 17 audio event tags.}
\begin{tabular}{lcc}
\hline
\textbf{Dev-set} & \textbf{F1} & \textbf{Error rate} \\
\hline
DCASE2017 baseline \cite{mesaros2017dcase} & 13.8  & 1.02 \\
\hline
Gated-CRNN-logMel & 47.20 & 0.76 \\
\hline
Fusion & \textbf{49.7} & \textbf{0.72} \\
\hline
\hline
\textbf{Eval-set} & \textbf{F1} & \textbf{Error rate} \\
\hline
DCASE2017 baseline \cite{mesaros2017dcase} & 28.4  & 0.93 \\
\hline
Gated-CRNN-logMel & 47.50 & 0.78 \\
\hline
Fusion & \textbf{51.8} & \textbf{0.73} \\
\hline
\end{tabular}%
	\label{tab:sed}
\end{table}

\section{Conclusions} \label{sec:conclusion}
In this paper, we proposed a unified method for audio tagging and weakly supervised sound event detection. Gated CRNN method is proposed, where the learnable gated linear units can help to select the most related features corresponding to the final labels. The temporal attention based localization method is also proposed to localize the occured events along the chunk in a weakly supervised mode. The final system puts us as in the 1st place with 57.7\% F1 score on the audio tagging sub-task of DCASE2017 challenge. We also ranks 2nd as a team in the SED sub-task. In the future, we will evaluate our proposed method on Audioset \cite{audioset}.

\vfill\pagebreak

\bibliographystyle{IEEEbib}
\bibliography{strings,refs}

\begin{thebibliography}{10}

\bibitem{kumar2016audio}
Anurag Kumar and Bhiksha Raj,
\newblock ``Audio event detection using weakly labeled data,''
\newblock in {\em Proceedings of ACM on Multimedia Conference}. ACM, 2016, pp.
  1038--1047.

\bibitem{bogdanov2013essentia}
Dmitry Bogdanov, Nicolas Wack, Emilia G{\'o}mez, Sankalp Gulati, Perfecto
  Herrera, Oscar Mayor, Gerard Roma, Justin Salamon, Jos{\'e}~R Zapata, Xavier
  Serra, et~al.,
\newblock ``Essentia: An audio analysis library for music information
  retrieval.,''
\newblock in {\em ISMIR}, 2013, pp. 493--498.

\bibitem{dimitrov2014analyzing}
Svilen Dimitrov, Jochen Britz, Boris Brandherm, and Jochen Frey,
\newblock ``Analyzing sounds of home environment for device recognition.,''
\newblock in {\em AmI}. Springer, 2014, pp. 1--16.

\bibitem{mesaros2016tut}
Annamaria Mesaros, Toni Heittola, and Tuomas Virtanen,
\newblock ``Tut database for acoustic scene classification and sound event
  detection,''
\newblock in {\em EUSIPCO}. IEEE, 2016, pp. 1128--1132.

\bibitem{mesaros2017dcase}
Annamaria Mesaros, Toni Heittola, Aleksandr Diment, Benjamin Elizalde, Ankit
  Shah, Emmanuel Vincent, Bhiksha Raj, and Tuomas Virtanen,
\newblock ``Dcase 2017 challenge setup: Tasks, datasets and baseline system,''
\newblock in {\em Proceedings of DCASE2017 Workshop}.

\bibitem{gemmeke2017audio}
Jort~F Gemmeke, Daniel~PW Ellis, Dylan Freedman, Aren Jansen, Wade Lawrence,
  R~Channing Moore, Manoj Plakal, and Marvin Ritter,
\newblock ``Audio set: An ontology and human-labeled dataset for audio
  events,''
\newblock in {\em ICASSP}, 2017.

\bibitem{xu2017trans}
Y~Xu, Q~Huang, W~Wang, , P~Foster, S~Sigtia, PJB Jackson, and MD~Plumbley,
\newblock ``Unsupervised feature learning based on deep models for
  environmental audio tagging,''
\newblock in {\em {IEEE}/{ACM} Trans. on audio, speech and language
  processing}, 2017.

\bibitem{qq2017icassp}
Q~Kong, Y~Xu, Wenwu Wang, and Mark~D Plumbley,
\newblock ``A joint detection-classification model for audio tagging of weakly
  labelled data,''
\newblock {\em ICASSP}, 2017.

\bibitem{ye2015acoustic}
Jiaxing Ye, Takumi Kobayashi, Masahiro Murakawa, and Tetsuya Higuchi,
\newblock ``Acoustic scene classification based on sound textures and events,''
\newblock in {\em Proceedings of ACM on Multimedia conference}. ACM, 2015, pp.
  1291--1294.

\bibitem{choi2016automatic}
Keunwoo Choi, George Fazekas, and Mark Sandler,
\newblock ``Automatic tagging using deep convolutional neural networks,''
\newblock {\em arXiv preprint arXiv:1606.00298}, 2016.

\bibitem{parascandolo2017convolutional}
Giambattista Parascandolo, Toni Heittola, Heikki Huttunen, Tuomas Virtanen,
  et~al.,
\newblock ``Convolutional recurrent neural networks for polyphonic sound event
  detection,''
\newblock {\em IEEE/ACM Transactions on Audio, Speech, and Language
  Processing}, vol. 25, no. 6, pp. 1291--1303, 2017.

\bibitem{yongIS2017}
Yong Xu, Qiuqiang Kong, Qiang Huang, Wenwu Wang, and Mark~D. Plumbley,
\newblock ``Attention and localization based on a deep convolutional recurrent
  model for weakly supervised audio tagging,''
\newblock in {\em INTERSPEECH}, 207, pp. 3083--3087.

\bibitem{dauphin2016language}
Yann~N Dauphin, Angela Fan, Michael Auli, and David Grangier,
\newblock ``Language modeling with gated convolutional networks,''
\newblock {\em arXiv preprint arXiv:1612.08083}, 2016.

\bibitem{nair2010rectified}
Vinod Nair and Geoffrey~E Hinton,
\newblock ``Rectified linear units improve restricted boltzmann machines,''
\newblock in {\em ICML}, 2010, pp. 807--814.

\bibitem{adavanne2017sound}
Sharath Adavanne, Pasi Pertil{\"a}, and Tuomas Virtanen,
\newblock ``Sound event detection using spatial features and convolutional
  recurrent neural network,''
\newblock {\em arXiv preprint arXiv:1706.02291}, 2017.

\bibitem{yongIJCNN2017}
Yong Xu, Qiuqiang Kong, Qiang Huang, Wenwu Wang, and Mark~D. Plumbley,
\newblock ``Convolutional gated recurrent neural network incorporating spatial
  features for audio tagging,''
\newblock in {\em IJCNN}, 2017, pp. 3461--3466.

\bibitem{kingma2014adam}
Diederik Kingma and Jimmy Ba,
\newblock ``Adam: {A} method for stochastic optimization,''
\newblock {\em arXiv preprint arXiv:1412.6980}, 2014.

\bibitem{hochreiter1997long}
Sepp Hochreiter and J{\"u}rgen Schmidhuber,
\newblock ``Long short-term memory,''
\newblock {\em Neural Computation}, vol. 9, no. 8, pp. 1735--1780, 1997.

\bibitem{chung2014empirical}
Junyoung Chung, Caglar Gulcehre, KyungHyun Cho, and Yoshua Bengio,
\newblock ``Empirical evaluation of gated recurrent neural networks on sequence
  modeling,''
\newblock {\em arXiv preprint arXiv:1412.3555}, 2014.

\bibitem{audioset}
Jort~F. Gemmeke, Daniel P.~W. Ellis, Dylan Freedman, Aren Jansen, Wade
  Lawrence, R.~Channing Moore, Manoj Plakal, and Marvin Ritter,
\newblock ``Audio set: An ontology and human-labeled dataset for audio
  events,''
\newblock in {\em ICASSP}, 2017.

\bibitem{Lee2017a}
Kyogu Lee, Donmoon Lee, Subin Lee, and Yoonchang Han,
\newblock ``Ensemble of convolutional neural networks for weakly-supervised
  sound event detection using multiple scale input,''
\newblock Tech. {R}ep., DCASE2017 Challenge, September 2017.

\bibitem{Chou2017}
Szu-Yu Chou, Jyh-Shing Jang, and Yi-Hsuan Yang,
\newblock ``{FrameCNN}: a weakly-supervised learning framework for frame-wise
  acoustic event detection and classification,''
\newblock Tech. {R}ep., DCASE2017 Challenge, 2017.

\end{thebibliography}

\end{document}